\documentstyle[11pt,AATS,epsf]{article}	

\begin{document}	

\title{Neutron-Capture Element Abundances and Cosmochronometry} 

\author{Christopher Sneden}
\affil{Department of Astronomy, University of Texas, Austin, TX 78712}

\author{John J. Cowan}
\affil{Department of Physics \& Astronomy, University of Oklahoma,
Norman OK 73019}

\author{Timothy C. Beers}
\affil{Department of Physics \& Astronomy, Michigan State University,
East Lansing, MI 48824}

\author{James W. Truran}
\affil{Department of Astronomy \& Astrophysics, University of Chicago,
Chicago, IL 60637}

\author{James E. Lawler}
\affil{Department of Physics, University of Wisconsin, Madison, WI 53706}

\author{George Fuller}
\affil {Department of Physics, University of California, San Diego,
La Jolla, CA 92093-0319}


\begin{abstract}
Abundance ratios of radioactive to stable neutron-capture elements in 
very metal-poor stars may be used to estimate the age of our Galaxy.  
But extracting accurate ages from these data requires continuing work 
on many fronts: a) identification of more low metallicity stars with 
neutron-capture element excesses; b) acquisition of the best high 
resolution stellar spectra; c) improvement in neutron-capture element
transition probabilities; d) calculation of more realistic nuclear 
models for, and interactions among the heaviest elements; and
e) and more self-consistent production predictions for these elements
in supernovae.  
This review discusses several of these areas and makes suggestions about
how to improve the accuracy of Galactic cosmochronometry.

\end{abstract}

\section{Introduction}

The dominant isotopes of elements with atomic numbers (Z~$>$~30)
are synthesized in neutron bombardment reactions during late stellar
evolution.
Some of the heaviest of these so-called neutron-capture ($n$-capture)
elements are radioactively unstable but long-lived on astrophysically 
interesting (many Gyr) time scales.
In principle, abundance analyses of old, metal-poor stars that compare
radioactive to stable $n$-capture elements can be used to determine the 
age of the Galactic halo.
But there are practical difficulties in applying this idea to 
most halo stars.
In this brief review we focus on studies of $n$-capture elements
relevant to cosmochronometry, discussing in turn general $n$-capture 
abundance trends with metallicity, detailed distributions of these 
elements in a few very well-observed stars, and possible new 
initiatives to improve the use of $n$-capture elements
in the description of early Galactic nucleosynthesis.

\section{Overall n-Capture Abundance Trends with Metallicity}

Abundances of $n$-capture elements vary respect to those of the Fe-peak 
by several orders of magnitude from star to star. 
This ``scatter'' is most apparent in the lowest metallicity regimes.
The onset metallicity of the $n$-capture scatter is poorly established,
but at least for stars with [Fe/H]~$<$~--2, the total range in 
[$<n$-capture$>$/Fe] is $\sim\pm$1.5~dex (e.g., Gilroy et~al. 1988, 
McWilliam et~al. 1995, Ryan et~al. 1996, Burris et~al. 2000).
(Here, [$<n$-capture$>$/Fe] stands for abundance ratios [Sr/Fe], [Ba/Fe], 
[La/Fe], etc.)
The abundance scatter is far greater than that ascribable to uncertainties in 
the observed stellar spectra, atomic data, or model atmosphere parameters. 
The variation is clearly seen in published spectra of very metal-poor stars
(e.g. Figure~16, McWilliam et~al. 1995; Figure~3, Burris et~al. 2000;
Figure~1, Westin et~al. 2000).
This provides direct evidence that local nucleosynthesis events added to 
an early Galactic halo ISM that remained poorly mixed on time scales 
corresponding to the formation of stars with metallicities as high as 
[Fe/H]~$\sim$~--2.

Abundance ratios among the $n$-capture elements in very metal-poor 
stars are distinctly non-solar.  
Spite \& Spite's (1979) high resolution spectroscopic investigation of
a few bright metal-poor stars provided the first convincing observational
evidence of this phenomenon.
In the metallicity domain 0~$>$~[Fe/H]~$>$ --2, the surveys cited above 
have shown that the ratio [Ba/Eu]~$\sim$~0, 
but by [Fe/H]~$\sim$~ --3 this ratio has declined to $\sim$--1.
Barium is synthesized most efficiently via slow neutron captures
(the $s$-process) while europium is predominantly created via rapid neutron
captures (the $r$-process).  
Thus observed low [Ba/Eu] values in the most metal-poor stars 
suggest (e.g. Truran 1981) that $r$-process products comprise most
of the $n$-capture abundances at lowest metallicities.
This supports the notion that $n$-capture element production in the
early Galaxy should have been from very short-lived, high mass stars
that generate large neutron fluxes during their death throes.
Products of the $s$-process come from longer-lived ($>$ $\sim$10$^8$~yr) 
low-to-intermediate mass stars.  
The evident lack of $s$-process contributions at [Fe/H]~$\sim$~--3\footnote{
Here we do not consider the so-called CH stars, ones that exhibit 
extremely large overabundances of C and of $n$-capture elements evidently 
created via the $s$-process (Norris et~al. 1997a,b; Hill et~al. 2000).  
Most such stars are binaries, undoubtedly victims of mass transfer from
higher-mass former AGB companions.  
These nucleosynthesis events eventually lead to the buildup of $s$-process
levels in the Galaxy, but contribute little at lowest metallicities.}
argues for a very rapid buildup of Galactic metallicity to this level
from first onset of star formation.

Less discussed is the lack of good correlation between abundances of
heavier (Z~$\geq$~56) and lighter (Z~=~38--40; Sr-Y-Zr)
$n$-capture elements. 
While the heavier elements appear to be $r$-process in origin, 
the lighter elements cannot be fit by the main $s$-process or $r$-process 
or their combination.
Instead,  it may require a complicated admixture of the {\it weak} 
$s$-process and the $r$-process to reproduce the Sr-Y-Zr abundance 
ratios in well-observed very metal-poor stars (e.g., Cowan et~al. 1995).
Moreover, Wasserburg et~al. (1996) have argued that the solar-system
$r$-process distribution results from two different sources or sites
-- one for the  lighter and one for the heavier $n$-capture elements.
It is also not yet fully understood why the abundances
of the Sr-Y-Zr group correlate much more closely with Fe-peak abundances
than do the heavier $n$-capture abundances,
although it might suggest a role for  the {\it weak} $s$-process
operating in massive stars early in the history of the Galaxy.
The lighter $n$-capture elements will be considered again in the next section.

\section{$n$-Capture Element Distributions in Well-Observed Halo Stars}

\begin{figure}[htb]	
\epsfxsize=25pc
\hspace*{0.3in}
\epsfbox{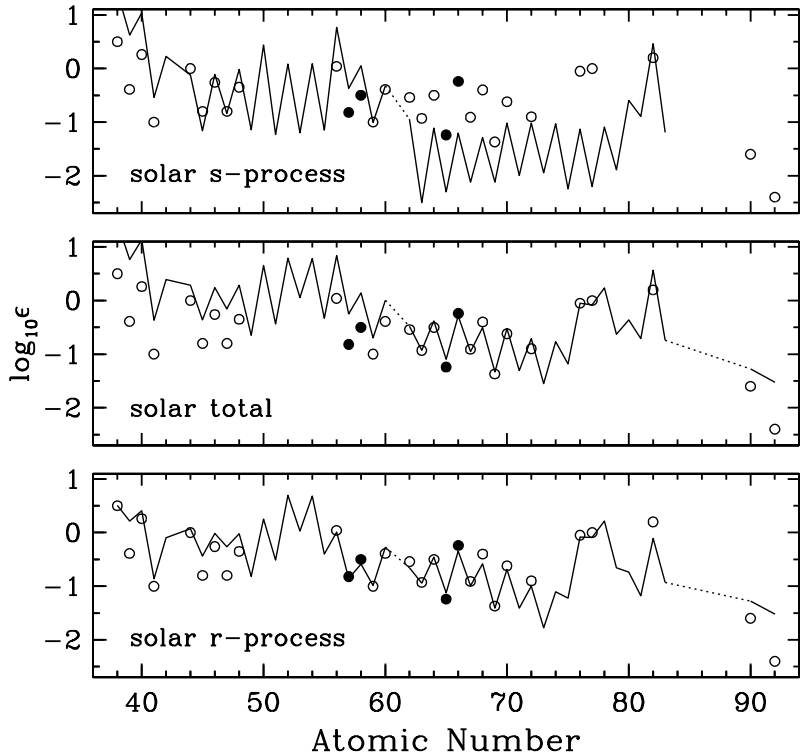}
\caption{CS~22892-052 $n$-capture abundances (points) and 
scaled solar system abundance distributions (solid and dashed lines).
Open circles are from Sneden et~al. (2000a), and filled circles are
abundances determined from the same observational data but with
new atomic data (see text).
The abundances units are
log$_{\rm 10}\epsilon$(A)~$\equiv$ 
log$_{\rm 10}$(N$_{\rm A}$/N$_{\rm H}$)~+~12.0.
The U (Z~=~92) abundance is an upper limit.
Uncertainties in individual abundances are $\sim\pm$0.05--0.10 dex.
The solar system distributions are from Burris et~al. (2000).
The scaling factors in the middle and bottom panels force agreement
with the Eu abundance; the scaling factor in the top panel is approximate,
for display purposes only.
\label{cs22fits}}
\end{figure}

To thoroughly examine the details of $n$-capture abundances in metal-poor
stars it is necessary to detect many elements over the entire Z~= 31--92
element range. 
This in turn requires identification of stars with 
[$<n$-capture$>$/Fe]~$\gg$~+0.5, in order to maximize the spectroscopic 
contrast between the strengths of $n$-capture and Fe-peak absorption features.
The few known extremely $n$-capture-rich metal-poor stars subjected to 
detailed abundance scrutiny have been serendipitous discoveries,
e.g. HD~115444 (Griffin et~al. 1982; Westin et~al. 2000), 
CS~22892-052 (McWilliam et~al. 1995; Sneden et~al. 2000a, and 
references therein), and now CS~31082-001 (Cayrel et~al. 2001).
Much attention has been given to CS~22892-052, and in 
Figure~\ref{cs22fits} we summarize the results of Sneden et~al. 
This figure also adds revised abundance determinations from the same 
observed spectra for elements with new transition probability and 
hyperfine structure data:  La (Lawler et~al. 2001a), Ce (Bi\'emont et~al.
2000), Tb (Lawler et al. 2001b), and Dy (Wickliffe et~al. 2000).
The CS~22892-052 abundances are compared with scaled solar-system 
$r$-process, $s$-process, and total abundance distributions.

Three features stand out in the plots of Figure~\ref{cs22fits}. 
First, the scaled solar $r$-process distribution almost perfectly fits
the observed abundances of the 18 heavier (Z~$>$~56) stable 
$n$-capture elements; the other two solar-system distributions fail
by large margins.
While the ``solar total'' curve appears to hold promise in comparison
to the observed abundances, normalizing the curve to Eu or Dy yields a 
mismatch by 0.5--0.9 dex in the Ba--Ce group.
Second, among the long-lived radioactive elements, the observed abundance 
of Th (Z~=~90) falls below the ``solar $r$-process'' curve, as does the 
observed upper limit to the abundance of U (Z~=~92).  
Thus $if$ these elements were originally synthesized in the same 
proportions with respect to the lighter stable $n$-capture elements that
occurred prior to solar system formation, then CS~22892-052 is substantially
older than the solar system, and in fact its $n$-capture elements were
synthesized $\sim$15~Gyr ago (e.g., Cowan et~al 1999).
Third, none of the solar system abundance distributions provide satisfactory
matches to the lighter (38~$\leq$~Z~$\leq$~48) $n$-capture element abundances
of this star.

The excellent agreement between observed abundances of the heavier stable 
elements and the ``solar $r$-process'' is not confined to CS~22892-052.
It is repeated in each of the few ultra-metal-poor stars ([Fe/H]~$<$~--2.5)
studied in comparable detail (Westin et~al. 2000; Johnson \& Bolte 2001).
Note in particular the halo giant BD+17~3248. 
This star has a nearly pure $r$-process signature among the $n$-capture 
elements, but with [Fe/H]~$\sim$ --2.0, it is the highest 
metallicity example of very $r$-process enriched material.
It will be of great interest to explore further the [Fe/H]~$>$~--2 domain
to discover at what metallicity the influence of individual $r$-process
synthesis events are lost in the general Galactic $n$-capture element
buildup.

Recently Cayrel et~al. (2001; see also their contribution to this 
conference volume) have discovered that the UMP star CS~31082-001 
([Fe/H]~$\sim$~--3) may have even larger $n$-capture element 
overabundances than does CS~22892-052.
Most interesting is their detection of $many$ transitions of \ion{Th}{2} and
the strongest \ion{U}{2} line; the combined abundances of two 
radioactive elements (with different half-lives) suggests to them that 
the ``age'' of the $n$-capture elements in this star is $\sim$13~Gyr.
Their ongoing study of this star finds an $observed$ [Th/Eu] ratio that 
is much larger than in CS~22892-052 (Sneden et~al. 2000a), HD~115444 
(Westin et~al. 2000), several other halo stars (Johnson \& Bolte 2001) 
and in globular cluster stars (Sneden et~al. 2000b).
This higher abundance ratio in CS~31082-001 would imply a young age 
comparable to the solar system, obviously inconsistent with this star's 
metallicity and halo membership.
This may indicate something unusual about the abundances in CS~31082-001,
or it may indicate that the production ratio of [Th/Eu] is not single-valued
and may lead to inconsistencies in age determinations 
(e.g., Goriely \& Clerbaux 1999).
It is clear that many more Eu and Th abundances need to be determined
for halo stars, in order to make some sense of this issue.

Finally, consider the non-conformity of CS~22892-052 abundances in the 
38~$\leq$~Z~$\leq$~48 domain to any of the scaled solar system curves.
Previous studies have indicated that the abundances of Sr--Zr did not fall 
on the same curve as the heavier elements, but this is the first 
case where elements between Z~= 40--50 have been detected. 
The difference  between the lighter and heavier $n$-capture element abundance 
distributions in CS~22892-052 (as shown in Figure~\ref{cs22fits}) has been
explained as the superposition of two distinct $r$-process events
(Wasserburg \& Qian 2000), or perhaps resulting from different regions
of the same neutron-rich jet of a core-collapse supernova (Cameron 2001)
that are responsible for synthesizing the lighter and heavier 
$n$-capture elements.
The breakpoint between the two r-process signatures is predicted to occur 
near Ba, as observed in CS~22892-052.
However, there has been no success thus far in fitting the individual 
abundances in this star, although the entire set of $n$-capture abundances 
in just CS~22892-052 has been used to constrain nuclear mass models and 
theoretical $r$-process predictions (Cowan et~al. 1999).
Further, since knowledge of the 41~$\leq$~Z~$\leq$~48 elements is confined 
so far to just CS~22892-052, what is most needed now are abundances of the 
lighter $n$-capture elements in more stars.

\section{Some Future Prospects}

\begin{figure}[htb]	
\epsfxsize=25pc
\hspace*{0.3in}
\epsfbox{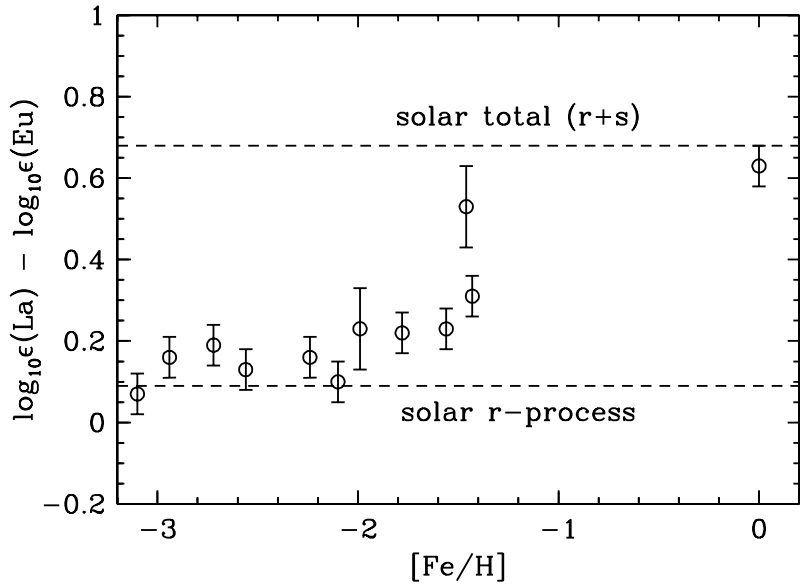}
\caption{Ratios of La to Eu abundances in a few representative stars
over most of the Galactic halo metallicity range.
The La abundances have been derived using new La II laboratory data
from Lawler et al. (2001a).
\label{laeu2}}
\end{figure}

The link between $n$-capture elemental abundances and Galactic time scales 
lies with the long-lived cosmochronometer elements.
So far, ratios of Th and U to each other and to other elements do not 
yield a consistent answer for the age of the Galactic halo when the 
observed abundances of a (very) few stars are compared
to zero-age predictions for them.
The [Th/Eu] ratios in CS~22892-052 and HD~115444 suggest ages of 
$\sim$15~Gyr, while that of CS~31082-001 yields $\sim$4~Gyr.
But the [Th/U] ratio of this star pushes the age back to $\sim$13~Gyr.
Johnson \& Bolte (2001) find that the ages of their five stars with
detected Th~II features can range from 11-15 Gyr, depending on assumptions
about initial Th/Eu production ratios.
Clearly both further theoretical (e.g. Cowan et~al. 2001) and
observational efforts are needed on this problem.
Extensive $n$-capture abundance distributions should be derived in at least 
$\sim$30 very low metallicity stars, to discover what is the normal abundance 
level of the radioactive chronometer elements in the Galactic halo.

Probably no other $absolute$ Galactic time scale information can be
deduced from $n$-capture elements in metal-poor stars. 
But recent improvements in stellar spectroscopic and laboratory atomic data 
permit renewed attack on an unsolved $relative$ time scale question: what is 
Galactic metallicity at which major contributions from $s$-process 
nucleosynthesis began to generally influence the halo ISM?
Most $s$-process synthesis is associated with the AGB phases of
intermediate-to-low mass stars, $\mathcal{M}$~$<$~8$\mathcal{M}_{\sun}$,
whose evolutionary time scales are $\ge$ $\sim$10$^8$~yr.
Metallicity regimes with little or no detectable $s$-process contributions 
are those resulting from the first waves of Galactic nucleosynthesis 
that happened on faster time scales.
The approximate [Fe/H] indicating a general rise in the $s$-process
should tell us how much the first nucleosynthesis burst contributed 
to overall Galactic metallicity.

This question has usually been empirically addressed by trying to find the
metallicity domain for substantial movement in [Ba/Eu] ratios from the
$r$-process-dominated value of $\sim$--0.9 at [Fe/H]~$\sim$~--3
toward the solar-system $r+s$ value of $\sim$0.0 that appears to be
complete by [Fe/H]~$\sim$~--1.5.
With such data, Gilroy et~al. (1988) suggested that the onset of major
Galactic $s$-process occurred at [Fe/H]~$\sim$~--2.3, while
Burris et~al. (2000) argue that this may happen at [Fe/H]~$\sim$~--2.8.
At present it is impossible to narrow this metallicity down to better than
somewhere in the range --3~$\leq$~[Fe/H]~$\leq$~--2. 

This situation may never change, because large observed star-to-star 
scatter exists in [Ba/Eu] ratios at all Galactic halo metallicities.
The scatter may be intrinsic to the stars, or it may simply be an artifact
of abundance analyses.
The problem lies in basic atomic structure. 
Unlike the first ions of neighboring rare earth elements, \ion{Ba}{2} has 
a structure that in cool stellar atmospheres gives rise to only five very 
strong transitions from lower energy levels in the visible spectral range.
All other \ion{Ba}{2} transitions are very weak ones from higher excitation
levels.
Significant hyperfine and isotopic splitting exists for the strong low
excitation lines, and Ba has 5 stable isotopes whose relative abundances
are synthesized in different proportions under the $r$- and $s$-processes;
estimates of these proportions becomes part of the abundance derivation
process (e.g. Magain 1995, Sneden et~al. 1996).
Abundances of Ba also are very sensitive to values of microturbulent
velocity assumed in the analyses.
All in all, it is difficult to believe [Ba/Eu] ratios in most metal-poor
stars to better than an uncertainty of $\sim\pm$0.2~dex, and this is
inadequate to map out the metallicity evolution of the $s$- and $r$-processes.

Fortunately other (Z~$\geq$~56) elements have very different abundances
resulting from the two $n$-capture events, and in particular [La/Eu] 
and [Ce/Eu] have $r$-/$s$-process sensitivities nearly equal to that
of [Ba/Eu] (e.g., see Figure~1 of Sneden et~al. 2001).
Both \ion{La}{2} and \ion{Ce}{2} have much more favorable atomic structures
than does \ion{Ba}{2}, and these elements have one very dominant isotope
each.
McWilliam (1997) advocated abandoning the use of [La/Eu] in favor of [Ba/Eu]
(see his Figure~10). 
But the data depicted in that diagram still has significant scatter.
Johnson \& Bolte (2001) argue from better data that observed [La/Eu]
ratios indicate $r$-process dominance in stars as metal-rich as 
[Fe/H]~$\sim$ --1.5.
Now, armed with higher resolution, higher S/N spectra and employing the 
new atomic data of Lawler et~al. (2001a), we are beginning
a large-sample survey of [La/Eu] in metal-poor stars.
Excellent line-by-line abundance agreement is seen for \ion{La}{2} 
in the Sun, CS~22892-052, and BD+17~3248 (Sneden et~al. 2001)
and in Figure~\ref{laeu2} we show the run of [La/Eu] with [Fe/H] for
a few very metal-poor but $n$-capture-rich stars.  
The small star-to-star scatter in [La/Eu] at lowest metallicities is so 
far very encouraging, but the sample is very small.
When this survey is completed we hope to be able to tell with far greater
certainty the metallicity at which the $s$-process makes substantial
contributions to most stars' $n$-capture abundances, and hence be
able to tell how fast the Galaxy may have increased its Fe-peak
metallicity before the deaths of the first intermediate-mass stars.

\acknowledgments  
Our work on $n$-capture elements in halo stars has been a collaborative
effort over many papers, and we appreciate our Co-authors for their
contributions over the years.
Emile Bi\'emont, Rica French, Jennifer Johnson, Jennifer Simmerer,
and Craig Wheeler are thanked for helpful discussions.
We are happy to acknowledge that this research has been supported by 
various NSF grants to the authors.



\begin{references}

\reference Burris, D. L., Pilachowski, C. A., Armandroff, T. A.,
Sneden, C., Cowan, J. J., \& Roe, H. 2000, \apj, 544, 302

\reference Cameron, A. G. W. 2001, Nuc. Phys. A, in press

\reference Cayrel, R., Hill, V., Beers, T. C., Barbuy, B., Spite, M., 
Spite, F., Plez, B., Andersen, J., Bonifacio, P., Francois, P., Molaro, P.,
Nordstrom, B., \& Primas, F. 2001, Nature, 409, 691

\reference Cowan, J. J., Burris, D. L., Sneden, C., McWilliam, A., 
\& Preston, G. W. 1995, \apj, 439, L51

\reference Cowan, J. J., McWilliam, A., Sneden, C., Burris, D. L.,
\& Preston, G. W. 1997, \apj, 480, 246

\reference Cowan, J. J., Pfeiffer, B., Kratz, K.-L., Thielemann, F.-K.,
Sneden, C., Burles, S., Tytler, D., \& Beers, T. C. 1999, \apj, 521, 194

\reference Cowan, J.J., Sneden, C., \& Truran, J. W. 2001, in Cosmic 
Evolution, ed. E. Vangioni-Flam and M. Cass\'e, (Singapore: World 
Scientific Publishing), in press

\reference Gilroy, K. K., Sneden, C., Pilachowski, C. A., \& Cowan, J. J.
1988, \apj, 327, 298

\reference Goriely, S., \& Clerbaux, B. 1999, \aap, 346, 798

\reference Griffin, R., Gustafsson, B., Viera, T., \& Griffin, R. 1982, \mnras,
198, 637

\reference Hill, V., Barbuy, B., Spite, M., Spite, F., Cayrel, R., Plez, B., 
Beers, T. C., Nordström, B., \& Nissen, P. E. 2000, \aap, 353, 557

\reference Johnson, J. A., \& Bolte, M. 2001, \apj, in press

\reference Lawler, J. E., Bonvallet, G., \& Sneden, C. 2001a, \apj, in press

\reference Lawler, J. E., Wickliffe, M. E., Cowley, C. R., \& Sneden, C. 
2001b, submitted

\reference Magain, P. 1995, \aap, 297, 696

\reference McWilliam, A. 1997, \araa, 35, 503

\reference McWilliam, A., Preston, G. W., Sneden, C., \& Searle, L.
1995, \aj, 109, 2757

\reference Norris, J. E., Ryan, S. G., \& Beers, T. C. 1997a, \apj, 488, 350

\reference Norris, J. E., Ryan, S. G., \& Beers, T. C. 1997b, \apj, 489, L169

\reference Palmeri, P., Quinet, P., Wyart, J.-F., \& Bi\'emont, E.
2000, Phys. Scr., 61, 323

\reference Ryan, S. G., Norris, J. E., \& Beers, T. C. 1996, \apj, 471, 254

\reference Sneden, C., Cowan, J.J., \& Truran, J. W. 2001, in 
Cosmic Evolution, ed. E. Vangioni-Flam and M. Cass\'e,
(Singapore: World Scientific Publishing), in press

\reference Sneden, C., Cowan, J. J., Ivans, I. I., Fuller, G. M., 
Burles, S., Beers, T. C., \& Lawler, J. E. 2000a, \apj, 533, L139

\reference Sneden, C., Johnson, J., Kraft, R. P., Smith, G. H., Cowan, J. J., 
Bolte, M. S. 2000b, \apj, 536, L85

\reference Sneden, C., McWilliam, A., Preston, G. W., Cowan, J. J.,
Burris, D. L., \& Armosky, B. J. 1996, \apj, 467, 819

\reference Spite, M., \& Spite, F., 1978, \aap, 67, 23

\reference Truran, J. W. 1981, \aap, 97, 391

\reference Wasserburg, G. J., Busso, M., \& Gallino, R. 1996, \apj, 466, 109

\reference Wasserburg, G. J., \& Qian, Y.-Z. 2000, \apj, 529, L21

\reference Westin, J., Sneden, C., Gustafsson, B., \& Cowan, J. J. 2000,
\apj, 530, 783

\reference Wickliffe, M. E., Lawler, J. E., \& Nave, G. 2000, JQSRT, 66, 363

\end{references}
\end{document}